\documentclass[conference,letterpaper]{IEEEtran}
\usepackage[utf8]{inputenc}
\usepackage{fancyhdr}
\usepackage{amsmath} 
\usepackage{amssymb} 
\usepackage{pifont} 
 \usepackage{cite}
\usepackage{url}
\usepackage{comment}
\usepackage{color, colortbl}
\usepackage{enumerate}
\usepackage{arcs}

\usepackage{fancyhdr}

\newtheorem{defini}{Definition}
\newtheorem{thm}{Theorem}
\newtheorem{exemple}{Example}
\newtheorem{pre}{Proposition}
\newtheorem{lem}{Lemma}
\definecolor{Gray}{gray}{0.9}
\begin{document}
\title{Authentication by Witness Functions}


\author{\IEEEauthorblockN{Jaouhar Fattahi$^1$ and Mohamed Mejri$^1$ and Emil Pricop$^2$}
\IEEEauthorblockA{$^1$Department of Computer  Science and Software  Engineering. Université Laval. Québec. Canada.\\ 
 \{jaouhar.fattahi.1@ulaval.ca ${\mid}$ mohamed.mejri@ift.ulaval.ca\}}
 \IEEEauthorblockA{$^2$Automatic Control, Computers and Electronics Department. Petroleum-Gas University of Ploiesti. Romania.\\ 
 \{emil.pricop@upg-ploiesti.ro\}}

}


%


\maketitle
\thispagestyle{plain}

\fancypagestyle{plain}{
\fancyhf{}	
\fancyfoot[L]{}
\fancyfoot[C]{}
\fancyfoot[R]{}
\renewcommand{\headrulewidth}{0pt}
\renewcommand{\footrulewidth}{0pt}
}

\pagestyle{fancy}{
\fancyhf{}
\fancyfoot[R]{}}
\renewcommand{\headrulewidth}{0pt}
\renewcommand{\footrulewidth}{0pt}

\begin{abstract}
Witness functions have recently been introduced in cryptographic protocols' literature as a new powerful way to prove  protocol correctness with respect to secrecy. In this paper, we extend them to the property of authentication. We show how to use them safely and we run an analysis on a modified version of the Woo-Lam protocol. We show that it is correct with respect to authentication.\\                                                                                                                                                                                                                                                                                                                                                                                                                                                                                                                                                                                                                                                                                                                                                                                                                                                                                                                                                                                                                                                                                                                                                                                                                                                                                                                                                                                                                                                                                                                                                                                                                                                                                                                                                                                                                                                                                                                                                                                                                                                                                                                                                                                                                                                                                                                                                                                                                                                                                                                                                                                                                                                                                                                                                                                                                                                                                                                                                                                                                                                                                                                                                                                                                                                                                                                                                                                                                                                                                                                                                                                                                                                                                                                                                                                                                                                                                                                                                                                                                                                                                                                                                                                                                                                                                                                                                                                                                                                                                                                                                                                                                                                                                                                                                                                                                                                                                                                                                                                                                                                                                                                                                                                                                                                                                                                                                                                                                                                                                                                                                                                                                                                                                                                                                                                                                                                                                                                                                                                                                                                                                                                                                                                                                                                                                                                                                                                                                                                                                                                                                                                                                                                                                                                                                                                                                                                                                                                                                                                                                                                                                                                                                                                                                                                                                                                                                                                                                                                                                                                                                                                                                                                                                                                                                                                                                                                                                                                                                                                                                                                                                                                                                                                                                                                                                                                                                                                                                                                                                                                                                                                                                                                                                                                                                                                                                                                                                                                                                                                                                                                                                                                                                                                                                                                                                                                                                                                                                                                                                                                                                                                                                                                                                                                                                                                                                                                                                                                                                                                                                                                                                                                                                                                                                                                                                                                                                                                                                                                                                                                                                                                                                                                                                                                                                                                                                                                                                                                                                                                                                                                                                                                                                                                                                                                                                                                                                                                                                                                                                                                                                                                                                                                                                                                                                                                                                                                                                                                                                                                                                                                                                                                                                                                                                                                                                                                                                                                                                                                                                                                                                                                                                                                                                                                                                                                                                                                                                                                                                                                                                                                                                                                                                                                                                                                                                                                                                                                                                                                                                                                                                                                                                                                                                                                                                                                                                                                                                                                                                                                                                                                                                                                                                                                                                                                                                                                                                                                                                                                                                                                                                                                                                                                                                                                                                                                                                                                                                                                                                                                                                                                                                                                                                                                                                                                                                                                                                                                                                                                                                                                                                                                                                                                                                                                                                                                                                                                                                                                                                                                                                                                                                                                                                                                                                                                                                                                                                                                                                                                                                                                                                                                                                                                                                                                                                                                                                                                                                                                                                                                                                                                                                                                                                                                                                                                                                                                                                                                                                                                                                                                                                                                                                                                                                                                                                                                                                                                                                                                                                                                                                                                                                                                                                                                                                                                                                                                                                                                                                                                                                                                                                                                                                                                                                                                                                                                                                                                                                                                                                                                                                                                                                                                                                                                                                                                                                                                                                                                                                                                                                                                                                                                                                                                                                                                                                                                                                                                                                                                                                                                                                                                                                                                                                                                                                                                                                                                                                                                                                                                                                                                                                                                                                                                                                                                                                                                                                                                                                                                                                                                                                                                                                                                                                                                                                                                                                                                                                                                                                                                                                                                                                                                                                                                                                                                                                                                                                                                                                                                                                                                                                                                                                                                                                                                                                                                                                                                                                                                                                                                                                                                                                                                                                                                                                                                                                                                                                                                                                                                                                                                                                                                                                                                                                                                                                                                                                                                                                                                                                                                                                                                                                                                                                                                                                                                                                                                                                                                                                                                                              
\end{abstract}


\section*{Notice}
 ${\mbox{\scriptsize{\copyright}}}$ \textit{2016 IEEE. Personal use of this material is permitted. Permission from IEEE must be obtained for all other uses, in any current or future media, including reprinting/republishing this material for advertising or promotional purposes, creating new collective works, for resale or redistribution to servers or lists, or reuse of any copyrighted component of this work in other works.}

%
\IEEEpeerreviewmaketitle

\section{Introduction}

Cryptographic protocols are distributed programs used everywhere. They are designed to provide security assurances in communications by using cryptographic primitives. Traditional assurances include secrecy, integrity and authentication, and more recently, anonymity, atomicity, verifiability, coercion-resistance, etc. It has been clear for a long time that cryptographic protocols are prone to serious design drifts.  Therefore, several formalisms~\cite{TrusComCortierK14} have been proposed  to check  protocols against flaws and vulnerabilities. In this paper, we focus on  authentication. Authentication is the property that should be satisfied by a protocol to be secure  against acceptance of a fraudulent  agent by ensuring that one party is who he claims to be. A protocol can reach authentication by means of secrecy when it succeeds  to provide evidence to one party that the second party manages to present a piece of data that could only have been generated by him (e.g. response to a challenge). This implies, very often, that some data can be unequivocally traced back to its origin. In some scenarios, authentication involves more than two parties where a trusted server could be used to introduce one party  $B$ to another party  $A$ and $A$ is assured that $B$ is trusted by the server in the required protocol semantics. It is worth mentioning that, of course, protocols can reach authentication with  no  need to secrecy, however, these protocols are not our target in this paper.

\section{Methodology}
Our way to prove that a protocol is correct with respect to authentication consists in reaching two intermediate purposes:
\begin{enumerate}
\item we prove that the secret used for authentication is  never disclosed to avoid any bad use of this latter by an opponent;
\item we make sure that the link between the authenticatee identity and the secret is never 	
untied. This is to guarantee that the secret emanates from the right origin.
\end{enumerate}
For that, we use the theory of witness functions~\cite{TheseJF,ChapterJFSpringer,WitnessArt1,Relaxed}. Witness  functions adequately estimate the level of security of each atomic message. Before this paper, we used to use them to prove  secrecy in a protocol by showing that it is increasing. That is to say, to show that the level of security of every single atomic message exchanged  in the protocol is always growing and never goes down between any two consecutive steps of receiving and sending. 
Our contribution here is the extension of witness functions for the authentication property. For that, we give sufficient conditions for authentication based on the two mighty bounds of a witness function such, if met, the protocol is automatically declared correct with respect to authentication. 

\begin{itemize}
\item in Sections \ref{sectionPreuveThFond} and \ref{sectionFonctionsetSelections}, we'll give a brief summary of the proof of correctness of increasing protocols when analyzed with safe functions and we give a guideline to build these functions;
\item in Section \ref{sectionWF},  we'll give a short overview on witness functions and we bring out their powerful bounds;
\item in Section \ref{suffCondAuth}, we'll give sufficient conditions to ensure protocols' correctness with respect to authentication based on witness functions;
\item in Section \ref{sectionAnWL2}, we'll analyze a modified version of the  Woo-Lam protocol;
\item finally, we'll compare our approach with similar ones and conclude. 
\end{itemize}

Please notice that all the notations we will use in this paper are given in the appendix. The reader is kindly requested to see them before moving forward.

\section{On the correctness of increasing protocols}\label{sectionPreuveThFond}                                                                                                                                                                                                                                                                                                                                                                                                                                                                                                                                                                                                                                                                                                                                                                                                                                                                                                                                                                                                                                                                                                                                                                                                                                                                                                                                                                                                                                                                                                                                                                                                                                                                                                                                                                                                                                                                                                                                                                                                                                                                                                                                                                                                                                                                                                                                                                                                                                                                                                                                                                                                                                                                                                                                                                                                                                                                                                                                                                                                                                                                                                                                                                                                                                                                                                                                                                                                                                                                                                                                                                                                                                                                                                                                                                                                                                                                                                                                                                                                                                                                                                                                                                                                                                                                                                                                                                                                                                                                                                                                                                                                                                                                                                                                                                                                                                                                                                                                                                                                                                                                                                                                                                                                                                                                                                                                                                                                                                                                                                                                                                                                                                                                                                                                                                                                                                                                                                                                                                                                                                                                                                                                                                                                                                                                                                                                                                                                                                                                                                                                                                                                                                                                                                                                                                                                                                                                                                                                                                                                                                                                                                                                                                                                                                                                                                                                                                                                                                                                                                                                                                                                                                                                                                                                                                                                                                                                                                                                                                                                                                                                                                                                                                                                                                                                                                                                                                                                                                                                                                                                                                                                                                                                                                                                                                                                                                                                                                                                                                                                                                                                                                                                                                                                                                                                                                                                                                                                                                                                                                                                                                                                                                                                                                                                                                                                                                                                                                                                                                                                                                                                                                                                                                                                                                                                                                                                                                                                                                                                                                                                                                                                                                                                                                                                                                                                                                                                                                                                                                                                                                                                                                                                                                                                                                                                                                                                                                                                                                                                                                                                                                                                                                                                                                                                                                                                                                                                                                                                                                                                                                                                                                                                                                                                                                                                                                                                                                                                                                                                                                                                                                                                                                                                                                                                                                                                                                                                                                                                                                                                                                                                                                                                                                                                                                                                                                                                                                                                                                                                                                                                                                                                                                                                                                                                                                                                                                                                                                                                                                                                                                                                                                                                                                                                                                                                                                                                                                                                                                                                                                                                                                                                                                                                                                                                                                                                                                                                                                                                                                                                                                                                                                                                                                                                                                                                                                                                                                                                                                                                                                                                                                                                                                                                                                                                                                                                                                                                                                                                                                                                                                                                                                                                                                                                                                                                                                                                                                                                                                                                                                                                                                                                                                                                                                                                                                                                                                                                                                                                                                                                                                                                                                                                                                                                                                                                                                                                                                                                                                                                                                                                                                                                                                                                                       

Hereafter, we remind an important result: ''A protocol keeps its secret inputs when analyzed with a safe function and shown increasing''.

\subsection{Safe functions}
\begin{defini}{(Well-formed Function)}\label{bienforme}
{
Let ${F}$ be a function and ${\mathcal{C}}$ be a context of verification. ${F}$ is ${\mathcal{C}}$-well-formed iff:
$\forall M,M_1,M_2 \subseteq {\mathcal{M}}, \forall \alpha \in {\mathcal{A}}({\mathcal{M}}) \mbox{:}$\\

\begin{tabular}{lll}
   ${F}(\alpha,\{\alpha\})$ & $=$ & $\bot$ \\
   ${F}(\alpha, {M}_1 \cup {M}_2)$ & $=$ & ${F}(\alpha, {M}_1)\sqcap{F}(\alpha,{M}_2)$ \\
   ${F}(\alpha,{M})$ & $=$ & $\top, \mbox{ if } \alpha \notin {\mathcal{A}}({M})$ \\
\end{tabular}
}$ $\\
\end{defini}

A well-formed function ${F}$ must return the infimum to an atom $\alpha$ that appears  clear in $M$ to say that  any participant knows it from $M$. It  returns to an atom in the union of two sets, the minimum of the two values calculated in each set separately. It returns the supremum to any atom $\alpha$ that does not appear in $M$ to say that nobody could get it from $M$.\\

\begin{defini}{(Full-Invariant-by-Opponent Function)}\label{spi}
{
Let ${F}$ be a function and ${\mathcal{C}}$ be a context of verification. 
${F}$ is full-invariant-by-Opponent iff: $\forall {M} \subseteq {\mathcal{M}}, m\in {\mathcal{M}},  \alpha \in {\mathcal{A}}(m)$:\\
$
 {M} \models_{\mathcal{C}} m \Rightarrow ({F}(\alpha,m) \sqsupseteq{F}(\alpha,{M})) \vee (\ulcorner K(I) \urcorner \sqsupseteq \ulcorner \alpha \urcorner).
$\\
}
\end{defini}

A full-invariant-by-opponent function ${F}$ is such, when it affects a security level to an atom $\alpha$ in a set of messages $M$, the opponent cannot deduce from this set some message $m$ in which this level decreases (i.e. ${F}(\alpha,m) \not \sqsupseteq{F}(\alpha,{M})$), unless he is expressly entitled  to know it (i.e. $\ulcorner K(I) \urcorner \sqsupseteq \ulcorner \alpha \urcorner$).\\

A function ${F}$ is s said to be safe if it is well-formed and full-invariant-by-Opponent.\\

\begin{defini}{(${F}$-Increasing Protocol)}\label{ProAbsCroi}
{
Let ${F}$ be a function and ${\mathcal{C}}$ be a context of verification and $p$ be a protocol.\\
$p$ is ${F}$-increasing iff: $\forall R.r \in R_G(p), \sigma \in \Gamma,  \alpha \in {\mathcal{A}}(r^+), \mbox{ we have: }{F}(\alpha, r^+\sigma)\sqsupseteq \ulcorner \alpha \urcorner \sqcap{F}(\alpha, R^-\sigma)$
}
\end{defini}
$ $\\
An ${F}$-increasing protocol must produce traces with atomic messages having always a security level, returned by ${F}$, higher when sending (i.e. in $ r^+\sigma$) than when receiving (i.e. in $R^-\sigma$) or than its level set in the context (i.e. $\ulcorner \alpha \urcorner$), if known. \\

\begin{thm}{(Sececy in Increasing Protocols)}\label{mainTh}
{
Let ${F}$ be a safe function and $p$ be an ${F}$-increasing protocol.
\begin{center}
$p$ keeps its secret inputs.
\end{center}
}
\end{thm}
$ $\\

Theorem \ref{mainTh} states that a protocol keeps its secret inputs when it is analyzed by a safe function $F$ and is shown increasing. This result is quite intuitive. In fact, if the opponent succeeds to obtain a secret $\alpha$ (clear) from the protocol then its security level given  by $F$ is the infimum since $F$ is well-formed. That cannot be due to the rules of the protocol  since this latter is $F\mbox{-increasing}$. That could not arise neither when the opponent uses his capabilities since $F$ is full-invariant-by-opponent. So, the secret is kept forever. The detailed proof is available in~\cite{Relaxed}.

\section{Guideline for Building Safe Functions}\label{sectionFonctionsetSelections}

In ~\cite{WitnessArt1} we propose a class  of safe selections: $S_{Gen}^{EK}$. Any selection $S$ in $S_{Gen}^{EK}$ must return to an atom $\alpha$ in a message $m$:

\begin{enumerate}
\item if $\alpha$ is encrypted by a key $k$ such that $k$ is the outer key that satisfies the condition $\ulcorner k^{-1} \urcorner \sqsupseteq \ulcorner \alpha \urcorner$ (we refer to it by the external protective key), a subset among $k^{-1}$ and the neighbors of $\alpha$ under the same protection by $k$;
\item for two messages linked by a function $f$ in $\Sigma$ other than an encryption by a protective key (e.g. pair), the union of two subselections in these two messages.
\item if $\alpha$ does not have a protective key in $m$, the infimum (all atoms);
\item if $\alpha$ does not appear in $m$, the supremum (the empty set);
\end{enumerate}


Among the selections of $S_{Gen}^{EK}$, we spotlight three practical ones:
\begin{enumerate}
\item the selection $S_{MAX}^{EK}:$ it returns to an atom $\alpha$ in a message $m$ encrypted by the external protective key $k$, all the principal identities inside the same protection by $k$, in addition to $k^{-1}$;
\item the selection $S_{EK}^{EK}:$ it returns to an atom $\alpha$ in a message $m$ encrypted by the external protective key $k$,  the key $k^{-1}$;
\item the selection $S_{N}^{EK}:$ it returns to an atom $\alpha$ in a message $m$ encrypted by the key external protective key $k$, all the principal identities inside the same protection by $k$;
\end{enumerate}
Any selection $S$ in $S_{Gen}^{EK}$ when composed to a adequate homomorphism $\psi$ returns a safe function $F=\psi \circ S$. 
We choose the homomorphism that returns for:
\begin{enumerate}
\item an agent, its identity;
\item the key $k^{-1}$, if selected, the set of agent identities  that are authorized to know it.
\end{enumerate}
We denote by $F_{MAX}^{EK}, F_{EK}^{EK}$ and $F_{N}^{EK}$ respectively the functions resulting from the compositions $\psi \circ S_{MAX}^{EK}, \psi \circ S_{EK}^{EK}$ and $\psi \circ S_{N}^{EK}$. We prove~\cite{WitnessArt1} that these functions are safe. In fact, since the selection for any atom $\alpha$  is performed in an invariant section protected by the external protective key $k$, then, to alter this section (to decrease the security level of  $\alpha$), the opponent must have priorly obtained the atomic key $k^{-1}$. At this point of the proof, his knowledge must satisfy: $\ulcorner K(I) \urcorner \sqsupseteq \ulcorner k^{-1} \urcorner$. Since the key $ k^{-1}$ satisfies: $\ulcorner k^{-1} \urcorner \sqsupseteq \ulcorner \alpha \urcorner$ then the knowledge of the opponent must satisfy $\ulcorner K(I) \urcorner \sqsupseteq \ulcorner \alpha \urcorner$ too seeing that the comparator "$\sqsupseteq$" is transitive. This is simply the definition of a full-invariant-by-opponent function. These functions are well-formed by construction, too. 

\begin{exemple}$ $\\

Context: $\ulcorner \alpha \urcorner=\{A, B, S\}$; $m=\{C.\{\alpha.D\}_{k_{as}}\}_{k_{ab}}$; ${k_{ab}^{-1}}={k_{ab}}, {k_{as}^{-1}}={k_{as}}$; $\ulcorner{k_{as}}\urcorner=\{A, S\}, \ulcorner{k_{ab}}\urcorner=\{A, B\}$. We have: \\
$S_{MAX}^{EK}(\alpha,m)=\{C, D, {k_{ab}^{-1}}\}$ and $F_{MAX}^{EK}(\alpha,m)=\psi\circ S_{MAX}^{EK}(\alpha,m)=\{C, D\}{ \cup}\ulcorner{k_{ab}^{-1}}\urcorner=\{C,D\} \cup \{A,B\}=\{A, B, C, D\}$.
\end{exemple} $ $\\


\section{Witness functions to eliminate the effect of variables}\label{sectionWF}

The functions  $F\in\{F_{MAX}^{EK}, F_{EK}^{EK}, F_{N}^{EK}\}$ we have  defined so far are not useful in practice since they operate on ground terms only. However, a static analysis should be run over messages of the generalized roles resulting from writing the protocol in a role-based specification~\cite{Debbabi11}. These messages contain variables that denote a content that an agent ignores and on which he cannot perform any verification. We give here a safe way to deal with variables. But now, let us  introduce derivation in Definition \ref{derivation}.$ $\\ 

\begin{defini}{[Derivation]}\label{derivation}
\begin{center}
\begin{tabular}{rrcll}
~~~~~~& $\partial_X \alpha$ & $=$ & $\alpha$ &\\
& $\partial_X X$ & $=$& $\epsilon$ &\\
& $\!\!\partial_X Y~$ & $=$ & $Y, X\not = Y$ &\\
& $\partial{[\overline{X}]} m$ & $=$ & $\partial_{\{{\mathcal{X}}_m\backslash \{X\}\}} m$ &\\
& $\partial_X f(m)$ & $=$ & $ {f}(\partial_X m), f\in \Sigma$ &\\
& $\partial_{S_1 \cup S_2}m$ & $=$& $\partial_{S_1}\partial_{S_2}m$&\\ \\
\end{tabular}
\end{center}
\end{defini}

Derivation simply eliminates variables from a message except a variable under evaluation. The expression $\partial m$ denotes a message $m$ after removing all variables in. The aim of derivation is to deprive  variables from playing any role when evaluating an atom in a message. The idea  is hence to apply $F$ on the derivative message instead of the message itself. Besides, any variable is evaluated as a block with no regard to its content (i.e quantified). In fact, the reason why we do not care about the content of a variable is that if a variable, globally evaluated, is shown increasing that means that it cannot be discovered by any unauthorized party, and consequently any content inside cannot be discovered by an unauthorized party, too. If it is not increasing, then the protocol will not satisfy our sufficient conditions and hence no decision with respect to secrecy is made on, thus, with respect to authentication, too. This way of treating variables allows us to  give any content of a variable (dynamically known) the same value of the variable itself (statically calculable). The way we evaluate an atom from within a derivative message is described by Definition \ref{Fder}. $ $ \\
\begin{defini}\label{Fder}
$ $\\
Let $m\in {\mathcal{M}}_p^{\mathcal{G}}$, $X \in {\mathcal{X}}_m$ and $m\sigma$ be a valid trace.\\
For all $\alpha \in {\mathcal{A}}(m\sigma)$, $\sigma\in\Gamma$, we denote by:
\[
F(\alpha, \partial [\overline{\alpha}] m\sigma) = \left\{
\begin{array}{ll}
F(\alpha,\partial m) & \mbox{if } \alpha \in {\mathcal{A}}(\partial m),\\
F(X,\partial [\overline{X}] m) & \mbox{if }\alpha \notin {\mathcal{A}}(\partial m) \\
& \mbox{and } \alpha =X\sigma. 


\end{array}
\right.
\]
\end{defini}

Since the expression $F(\alpha, \partial [\overline{\alpha}] m\sigma)$ does not depend on substitution (i.e the run $\sigma$), we denote it simply by $F'(\alpha,m)$. Although the derivative function $F'$ allows us to neutralize variables, which is as such an important intermediate result, using it as is to analyze protocols is not safe because it is not even a function in the protocol. In fact, when we want to evaluate the security level of $\alpha$ in the trace $\{\alpha.A.B\}_{k_{cd}}­$ for example, we must return to the generalized roles to see from which message (with variables) this trace is generated and we may realize that more than one message are able to generate it. For instance, the messages $m_1=\{\alpha.A.X\}_{k_{cd}}­$ and $m_2=\{\alpha.Y.B\}_{k_{cd}}­$ are both candidates to generate the trace $\{\alpha.A.B\}_{k_{cd}}­$ (i.e. possible sources of it). If we refer to the first source (i.e. to $m_1$), the security level of $\alpha$ returned by $F'$ (with $F=F_{MAX}^{EK}$) is: 
$$F'(\alpha,\{\alpha.A.X\}_{k_{cd}})=F(\alpha,\{\alpha.\textcolor{red}{A}\}_{k_{\textcolor{red}{cd}}})=\{A, C, D\}$$
Whereas, if we refer to the second source (i.e. to $m_2$), its security level  is: 
$$F'(\alpha,\{\alpha.Y.B\}_{k_{cd}})=F(\alpha,\{\alpha.\textcolor{magenta}{B}\}_{k_{\textcolor{magenta}{cd}}})=\{B, C, D\}$$
Therefore, we cannot rely on $F'$ to analyze a protocol. To overcome this insufficiency, we define the witness functions. A witness function  looks for  all the sources of a ground term $m\sigma$ in the finite set  ${\mathcal{M}}_p^{\mathcal{G}}$, applies $F'$ to all of them, and returns the minimum. This minimum must satisfy both existence and uniqueness in the security lattice. Since a secret is always evaluated under the protection of a protective key, the research  of the  sources of a closed message in ${\mathcal{M}}_p^{\mathcal{G}}$ is then restricted to encryption patterns. We denote by  $\tilde{\mathcal{M}}_p^{\mathcal{G}}$ the set of all  encryption patterns generated by the protocol (renamed).\\

\begin{defini}{[witness function]}\label{WF}
Let $m\in {\mathcal{M}}_p^{\mathcal{G}}$, $X \in {\mathcal{X}}_m$ and $m\sigma$ be a valid trace.
Let $p$ be a protocol and $F$ be a safe function. 
We define a witness function ${\mathcal{W}}_{p,F}$ for all $\alpha \in {\mathcal{A}}(m\sigma)$, $\sigma\in\Gamma$, as follows: 
\[{{{\mathcal{W}}}}_{p,F}(\alpha,m\sigma)=\underset{{\{(m',\sigma') \in \tilde{\mathcal{M}}_p^{\mathcal{G}}\otimes\Gamma|m'\sigma' = m \sigma \}}}{\sqcap} \!\!\!\!\!\!\!\!\!\!\!\!\!\!\!\!\!\!\!\!\!\!\!\!\!\!\!F'(\alpha, m'\sigma')\]
\end{defini}


Using a witness function as is to analyze a protocol is a very tedious  process since we cannot enumerate all the valid traces $m\sigma$ and their sources statically. For that we bind it into two bounds that do not depend on substitution (i.e on $\sigma$). The upper bound of a witness function returns a minimum set of confirmed principal identities for any $\alpha$ in $m$ whereas the lower bound returns the set of all principal identities from all the possible sources of $m$ (that are unifiable with it) including the odd ones.  The lower bound hunts so any odd principal identity and interprets it as an attack. The proof is simple since we have always  $\{(m',\sigma') \in \tilde{\mathcal{M}}_p^{\mathcal{G}}\otimes\Gamma|m'\sigma' = m \sigma \} \subseteq \{(m',\sigma') \in \tilde{\mathcal{M}}_p^{\mathcal{G}}\otimes\Gamma|m'\sigma' = m \sigma' \}$ in the security lattice whatever $\sigma$. This is expressed by Proposition~\ref{prePAT}.\\

\begin{pre}{[binding a witness function]}\label{prePAT}
Let $m\in{\mathcal{M}}_p^{\mathcal{G}}$. Let ${\mathcal{W}}_{p,F}$ be a witness function.
For all $\sigma \in \Gamma$ we have:
$$F'(\alpha, m)\sqsupseteq {{{\mathcal{W}}}}_{p,F}(\alpha,m\sigma)\sqsupseteq \!\!\!\! \!\!\!\! \!\!\! \underset{{\{(m',\sigma') \in \tilde{\mathcal{M}}_p^{\mathcal{G}}\otimes\Gamma|m'\sigma' = m \sigma' \}}}{\sqcap} \!\!\!\!\!\!\!\!\!\!\!\!\!\!\!\!\!\!\!\!\!\!\!\!\!\!\!\!F'(\alpha, m'\sigma')$$
\end{pre}

\section{Sufficient conditions for authentication (Contribution)} \label{suffCondAuth}

The Lemma \ref{PAT} declares a decision criterion for secrecy  based on the bounds  of a witness function. The upper bound returns the set of original and trusted identities only from the received message. The lower bound returns the set of all the identities including those that could be inserted by a masquerader by substituting a regular message when sent. The criterion makes sure that no odd identity could be inserted in the evaluation neighborhood of a given  atom in the sent message.  The  proof  of Lemma \ref{PAT} results directly from Proposition \ref{prePAT} and Theorem \ref{mainTh}. \\

\begin{lem}{[Decision for Secrecy]}\label{PAT}
Let $p$ be a protocol. Let ${\mathcal{W}}_{p,F}$ be a witness function.
$p$ is correct with respect to secrecy if:
$\forall R.r \in R_G(p), \forall \alpha \in {\mathcal{A}}{(r^+ )}$ we have:
$$\underset{{\{(m',\sigma') \in \tilde{\mathcal{M}}_p^{\mathcal{G}}\otimes\Gamma|m'\sigma' = r^+ \sigma' \}}}{\sqcap} \!\!\!\!\!\!\!\!\!\!\!\!\!\!\!\!\!\!\!\!\!\!\!\!\!\!\!\!\!F'(\alpha,  m'\sigma') \sqsupseteq \ulcorner \alpha \urcorner \sqcap F'(\alpha, R^-)$$
\end{lem}

Theorem \ref{PAT2} forwards two conditions such, if met, a protocol is automatically declared correct for authentication. The first one ensures that the protocol does not disclose its secret inputs. An immediate outcome of this first result is that the message the verifier last receives for authentication is in a safe state and had not been subverted by a masquerader. The second one makes sure that the binding between the challenge and the identity of the identifier is not broken.  Therefore, we call once again the upper bound of the witness function to confirm the originality of the claimer identity by evaluating the challenge in the  message that should authenticate the claimer to the verifier. The second clause of the second condition is trivial  and introduced just to make sure that the challenge is not received in a public state.\\

\begin{thm}{[Decision for Authentication]}\label{PAT2}
Let $p$ be a protocol. Let ${\mathcal{W}}_{p,F}$ be a witness function.
$p$ is correct with respect to authentication if:
\begin{enumerate}
\item Lemma \ref{PAT} is satisfied;
\item Let $\alpha$ be the challenge in a message $m$ received by a verifier $B$ to authenticate an agent $A$. We  have:
$$A \in F'(\alpha, m)  \wedge F'(\alpha,  m)  \sqsupset \bot$$
\end{enumerate}
\end{thm}
It is worth mentioning  that:
\begin{itemize}
 
\item  the conditions set by Theorem \ref{PAT2} can be verified statically as they use the two bounds of a witness function only and  both of them are statically determinable;
\item their verification is linear under the assumption of perfect encryption since the most costly operation is unification in the lower bound of a witness function which is linear under that assumption. However, under nonempty equational theories, it will vary from one to another. In a future work, we will give new results related to this question.
   
\end{itemize}
\section{Analysis of a modified version of the Woo-Lam Protocol  with a witness function for authentication  (Contribution)} \label{sectionAnWL2}

The original version of the Woo-Lam protocol has been proven incorrect  by several means~\cite{ChaoMejri04,ShaikTrustCOM,avispa1}. Hereafter, we analyze an modified version of this protocol with a witness function and we prove that it is correct with respect to authentication. This version  is  denoted by $p$ in Table~\ref{WLMV:protv2p}.

\begin{center}
          \begin{table}[h]
      \begin{center}    
         \caption{The Woo-Lam Protocol (modified version)}
         \label{WLMV:protv2p} 
               $\begin{array}{|l|}
               \hline\\
                   \begin{array}{llll}
                    p   :=&\langle 1,A\rightarrow B: A\rangle.  \\
                    & \langle 2,B\rightarrow A: N_b\rangle. \\
                    & \langle 3,A\rightarrow B: \{ B.k_{ab}\}_{k_{as}}\rangle. \\
                    & \langle 4,B\rightarrow S:\{A.N_b.\{B.k_{ab}\}_{k_{as}}\}_{k_{bs}}\rangle.  \\
                    & \langle 5,S\rightarrow B:\{N_b.\{A.k_{ab}\}_{k_{bs}} \}_{k_{bs}}\rangle\\ &\\
                    \end{array} \\ \hline  \end{array}$
\end{center}
         \end{table}
\end{center}

    The   role-based   specification  of $p$  is ${\cal R}_G(p) = \{{\cal A}_G ^1,
    ~{\cal  A}_G  ^2,~{\cal B}_G ^1,~{\cal B}_G ^2,~{\cal B}_G ^3,~ {\cal S}_G ^1\}$,
    where the generalized roles ${\cal A}_G ^1$, ${\cal A}_G ^2$ of $A$ are as follows:
    \[\begin{array}{l}
            \begin{array}{lllll}
                    {\cal A}_G ^1 =& \langle  i.1, A    & \rightarrow & I(B):&  A
                    \rangle\\
            \end{array}\\
                    \\
             \begin{array}{llllll}
                    {\cal A}_G ^2=& \langle  i.1,  A    & \rightarrow & I(B):&  A \rangle .\\
                    & \langle i.2,  I(B) & \rightarrow & A:&  X \rangle .\\
                    & \langle i.3,  A    & \rightarrow & I(B):&  \{B.k_{ab}^i\}_{k_{as}}\rangle
             \end{array}\end{array}\]

     The generalized roles ${\cal B}_G ^1$, ${\cal B}_G ^2$, ${\cal B}_G
    ^3$  of $B$ are as follows:
            \[\begin{array}{l}
            \begin{array}{lllll}
                {\cal B}_G ^1=& \langle i.1,  I(A) & \rightarrow & B:&  A \rangle .\\
                        & \langle i.2,  B    & \rightarrow & I(A) :&  N_b^i \rangle  \\
            \end{array}\\ \\
            \begin{array}{lllll}
                    {\cal B}_G ^2=& \langle i.1,  I(A) & \rightarrow & B :&  A \rangle .\\
                    & \langle i.2,  B    & \rightarrow & I(A):&  N_b ^i\rangle .\\
                    & \langle i.3,  I(A) & \rightarrow & B:&  Y \rangle .\\
                    & \langle i.4,  B    & \rightarrow & I(S):& \{A.N_b^i.Y \}_{k_{bs}}\rangle
            \end{array}\\ \\
            \begin{array}{lllll}
                    {\cal B}_G ^3=& \langle i.1,  I(A) & \rightarrow & B:&  A \rangle .\\
                    & \langle i.2,  B    & \rightarrow & I(A):&  N_b ^i\rangle.\\
                    & \langle i.3,  I(A) & \rightarrow & B:&  Y \rangle.\\
                    & \langle i.4, B    & \rightarrow & I(S):& \{A.N_b^i.Y \}_{k_{bs}}\rangle.  \\
                    & \langle i.5,  I(S) & \rightarrow & B:&  \{N_b^i.\{A.Z\}_{k_{bs}}\}_{k_{bs}}\rangle \\
            \end{array}
             \end{array}\]

             The  generalized role $ {\cal S}_G ^1$ of $S$ is as follows:
            \[\begin{array}{l}
             \begin{array}{lllll}
                            {\cal S}_G ^1= & \langle i.4, I(B) & \rightarrow & S:& \{A.U.\{B.V\}_{k_{as}}
                            \}_{k_{bs}} \rangle. \\
                                    & \langle i.5,S & \rightarrow & I(B):& \{U.\{A.V\}_{k_{bs}}\}_{k_{bs}}\rangle\\
               \end{array}
            \end{array} \]
            
Let us have a context of verification such that: \\

$\ulcorner k_{as}\urcorner=\{A, S\}$; $\ulcorner k_{bs}\urcorner=\{B, S\}$; $\ulcorner k_{ab}^i\urcorner=\{A, B, S\}$; \\

$\ulcorner N_b^i \urcorner=\bot$; $\forall A\in {\cal{I}}, \ulcorner A \urcorner=\bot$.\\


Let $F= F_{MAX}^{EK}$; ${\mathcal{W}}_{p,F}= {\mathcal{W}}_{{p},F_{MAX}^{EK}}$;\\

We denote  by ${\Upsilon}_{p,F}(\alpha,m)$ the lower bound $\underset{\{{(m',\sigma') \in \tilde{\cal{M}}_{p}^{\cal{G}}}\otimes \Gamma{|m'\sigma' = m\sigma' \}}}{\sqcap}\!\!\!\!\!\!\!\!\!\!  F'(\alpha, m'\sigma')$ of the witness function ${\mathcal{W}}_{p,F}(\alpha,m)$.\\

The set of messages generated by $p$ is  ${\cal{M}}_{p}^{\cal{G}}=\{
A_1,
X_1,
\{B_1.K_{A_{2}B_{1}}^i\}_{K_{A_{2}S_{1}}},
A_3 ,
N_{B_{2}}^i,
Y_1,$ $
\{A_4.N_{B_{3}}^i.Y_2 \}_{K_{{B_{3}S_2}}},
\{N_{B_{4}}^i.\{A_5.Z_1\}_{K_{B_{4}S_3}}\}_{K_{B_{4}S_3}},\\
\{A_6.U_1.\{B_5.V_1\}_{K_{{A_6}S_4}}\}_{K_{{B_5}S_4}},
\{U_2.\{A_7.V_2\}_{K_{B_{6}S_5}}\}_{K_{B_{6}S_5}}
 \}$\\
 
 The variables are denoted by $X_1, Y_2, Z_1, U_1, U_2, V_1$ and $V_2$;\\ 
 
After eliminating   duplicates in ${\cal{M}}_{p}^{\cal{G}}$ and the messages that are not encryption patterns, we have:\\ $ $\\
$\tilde{\cal{M}}_{p}^{\cal{G}}= \{
\{B_1.K_{A_{2}B_{1}}^i\}_{K_{A_{2}S_{1}}},
\{A_4.N_{B_{3}}^i.Y_2 \}_{K_{{B_{3}S_2}}},$ $
\{N_{B_{4}}^i.\{A_5.Z_1\}_{K_{B_{4}S_3}}\}_{K_{B_{4}S_3}},
\{A_6.U_1.\{B_5.V_1\}_{K_{{A_6}S_4}}\}_{K_{{B_5}S_4}},\\
\{U_2.\{A_7.V_2\}_{K_{B_{6}S_5}}\}_{K_{B_{6}S_5}}
 \}$\\

\subsection{Analysis of the Generalized Roles of $A$}

According to the generalized role of  $A$, an agent $A$ may take part in some session $S^{i}$  in which he receives an unkown message $X$ and sends the message $\{B.k_{ab}^i\}_{k_{as}}$. This is described by the following rule: \[{S^{i}}:\frac{X}{\{B.k_{ab}^i\}_{k_{as}}}\]

\noindent{-Analysis of the messages exchanged  in $S^{i}$:}\\
\\
1- For $k_{ab}^i$:\\

a- Receiving step: $R_{S^{i}}^-=X$ \textit{(when receiving, we use the upper bound)}

\begin{equation}
F'(k_{ab}^i, R_{S^{i}}^-)=F(k_{ab}^i,\partial  [\overline{k_{ab}^i}]X )=F(k_{ab}^i,\epsilon)=\top
\label{eq1}
\end{equation}

b- Sending step: $r_{S^{i}}^+=\{B.k_{ab}^i\}_{k_{as}}$ \textit{(when sending , we use the lower bound)}\\

 $\forall k_{ab}^i.\{(m',\sigma') \in \tilde{\cal{M}}_{p}^{\cal{G}}\otimes \Gamma|{m'\sigma' = r_{S^{i}}^+\sigma' } \}$\\
 
 $=\forall k_{ab}^i.\{(m'\sigma') \in \tilde{\cal{M}}_{p}^{\cal{G}}\otimes \Gamma|{m'\sigma' = \{B.k_{ab}^i\}_{k_{as}}\sigma' } \}$ \\
 
 $=\{(\{B_1.K_{A_{2}B_{1}}^i\}_{K_{A_{2}S_{1}}},\sigma_1')\}$ such that: $$ \sigma_1'=\{ B_1 \longmapsto B, K_{A_{2}B_{1}}^i \longmapsto k_{ab}^i, {K_{A_{2}S_{1}}} \longmapsto {k_{as}}\}$$
 
${\Upsilon}_{p,F}(k_{ab}^i,\{B.k_{ab}^i\}_{k_{as}})$\\

$=\{\mbox{Definition of the lower bound of the witness function}\}$\\

$F'(k_{ab}^i, \{B_1.K_{A_{2}B_{1}}^i\}_{K_{A_{2}S_{1}}} \sigma_{1}')$\\

$=\{\mbox{Setting the static neighborhood}\}$\\

$F'(k_{ab}^i, \{B.k_{ab}^i\}_{k_{as}} \sigma_{1}')$ \\

$=\{\mbox{Definition } \ref{Fder}\}$
\\

$F(k_{ab}^i,\partial[\overline{k_{ab}^i}] \{B.k_{ab}^i\}_{k_{as}})$ \\

$=\{\mbox{Derivation in Definition } \ref{derivation}\}$
\\

$F(k_{ab}^i,\{B.k_{ab}^i\}_{k_{as}})$ \\

$=\{\mbox{Since } F=F_{MAX}^{EK}\}$\\

$\{B, A, S\}$\\

Then, we have:

\begin{equation}
{\Upsilon}_{p,F}(k_{ab}^i,\{B.k_{ab}^i\}_{k_{as}})=\{B, A, S\}
\label{eq2}
\end{equation}
\\
2- Conformity  with Lemma \ref{PAT}:\\ 

From (\ref{eq1}) and (\ref{eq2}) and since $ \ulcorner k_{ab}^i \urcorner=\{B, A, S\}$ in the context, we have:
\begin{equation}
{\Upsilon}_{p,F}(k_{ab}^i,\{B.k_{ab}^i\}_{k_{as}}) \sqsupseteq \ulcorner k_{ab}^i \urcorner \sqcap F'(k_{ab}^i, X )
\label{eq3}
\end{equation}

Then,  the generalized role of $A$ respects Lemma \ref{PAT}. ~~~~ (I)

\subsection{Analysis of the generalized roles of $B$}
According to the generalized role of $B$, an agent $B$ may take part in two subsequent sessions: $S^{i}$ and $S^{j}$ such that $j>i$. In the first session $S^{i}$, the agent $B$ receives the identity $A$ and sends the nonce $N_b ^i$. In the second one $S^{j}$, he receives an unknown message $Y$ and he sends the message $\{A.N_b^i.Y \}_{k_{bs}}$. This is described by the following rules:

\[{S^{i}}:\frac{A}{N_b ^i} ~~~~~~~~~~~~~~~~~~~~~~~~ {S^{j}}:\frac{Y}{\{A.N_b^i.Y \}_{k_{bs}}}\]
$ $\\
\noindent{B.1- Analysis of the messages exchanged  in  $S^{i}$:}\\
\\
1- For $N_b^i$:\\

Since $N_b^i$ is set public in the context (i.e. $\ulcorner N_b^i \urcorner=\bot$), then, we have directly:
\begin{equation}
{\Upsilon}_{p,F}(N_b^i,N_b^i)\sqsupseteq \ulcorner N_b^i \urcorner \sqcap F'(N_b^i,A )\\
\label{eq4}
\end{equation}
$ $\\
\noindent{B.2- Analysis of the messages exchanged  in $S^{j}$:}\\
$ $\\
1- For $N_b^i$:\\

Since $N_b^i$ is set public in the context (i.e. $\ulcorner N_b^i \urcorner=\bot$), then, we have directly:
\begin{equation}
{\Upsilon}_{p,F}(N_b^i,\{A.N_b^i.Y \}_{k_{bs}})\sqsupseteq \ulcorner N_b^i \urcorner \sqcap  F'(N_b^i,Y )\\
\label{eq5}
\end{equation}
2- $\forall Y$:\\

Since when receiving, we have: $$F'(Y,Y )=F(Y,\partial  [\overline{Y}]Y )=F(Y,Y)=\bot$$

Then, we have directly:
\begin{equation}
{\Upsilon}_{p,F}(Y,\{A.N_b^i.Y \}_{k_{bs}})\sqsupseteq \ulcorner Y \urcorner \sqcap F'(Y,Y )\\
\label{eq6}
\end{equation}

3- Conformity  with Lemma \ref{PAT}:\\

From (\ref{eq4}), (\ref{eq5}) and (\ref{eq6}), we have:  the generalized role of $B$ respects Lemma \ref{PAT}. ~~~~~~~~~~~~~~~~~~~~~~~~~~~~~~~~~~~~~~~~~~~~~~~~~~~~~(II)\\

\subsection{Analysis of the generalized roles of $S$}$ $\\

According to the generalized role of  $S$, an agent $S$ may take part in some session $S^{i}$  in which he receives the message $ \{A.U.\{B.V\}_{k_{as}}\}_{k_{bs}}$ and sends the message $ \{U.\{A.V\}_{k_{bs}}\}_{k_{bs}}$. This is described by the following rule: \[{S^{i}}:\frac{\{A.U.\{B.V\}_{k_{as}}\}_{k_{bs}}}{ \{U.\{A.V\}_{k_{bs}}\}_{k_{bs}}}\]

1- $\forall U$:\\

a- Receiving step: $R_{S^{i}}^-=\{A.U.\{B.V\}_{k_{as}}\}_{k_{bs}}$ \textit{(when receiving, we use the upper bound)}

\begin{equation} 
\begin{tabular}{lll}
   $\!\!\!\!F'(U,\{A.U.\{B.V\}_{k_{as}}\}_{k_{bs}} )$ & $\!\!\!\!=$ & $\!\!\!\!F(U,\partial  [\overline{U}]\{A.U.\{B.V\}_{k_{as}}\}_{k_{bs}} )$  \\
   & $\!\!\!\!=$ & $\!\!\!\! F(U,\{A.U.\{B\}_{k_{as}}\}_{k_{bs}} )$ \\
    & $\!\!\!\!=$  & $\!\!\!\!\{A, B, S\}$ \\
\end{tabular}
\label{eq7}
\end{equation}

b- Sending step: $r_{S^{i}}^+=\{U.\{A.V\}_{k_{bs}}\}_{k_{bs}}$ \textit{(when sending , we use the lower bound)}\\

 $\forall U.\{(m',\sigma') \in \tilde{\cal{M}}_{p}^{\cal{G}}\otimes\Gamma|{m'\sigma' = r_{S^{i}}^+\sigma' } \}$\\
 
 $=\forall U.\{(m',\sigma') \in \tilde{\cal{M}}_{p}^{\cal{G}}\otimes\Gamma|{m'\sigma' = \{U.\{A.V\}_{k_{bs}}\}_{k_{bs}} \sigma' } \}$ \\
 
 $=\{(\{\{U_2.\{A_7.V_2\}_{K_{B_{6}S_5}}\}_{K_{B_{6}S_5}},\sigma_1')\}$ such that: 
 
 $$\sigma_1'=\{ U_2 \longmapsto U, A_7\longmapsto A, V_2 \longmapsto V, {K_{B_{6}S_5}}\longmapsto  {k_{bs}}\}$$\\

${\Upsilon}_{p,F}(U,\{U.\{A.V\}_{k_{bs}}\}_{k_{bs}})$\\

$=\{\mbox{Definition of the lower bound of the witness function}\}$\\

$F'(U,\{U_2.\{A_7.V_2\}_{K_{B_{6}S_5}}\}_{K_{B_{6}S_5}} \sigma_{1}') $\\

$=\{\mbox{Setting the static neighborhood}\}$\\

$F'(U, \{U_2.\{A.V\}_{k_{bs}}\}_{k_{bs}} \sigma_{1}')$ \\

$=\{\mbox{Definition } \ref{Fder}\}$
\\

$F(U_2,\partial[\overline{U_2}] \{U_2.\{A.V\}_{k_{bs}}\}_{k_{bs}})$ \\

$=\{\mbox{Derivation in Definition } \ref{derivation}\}$
\\

$F(U_2,\{U_2.\{A\}_{k_{bs}}\}_{k_{bs}})$ \\

$=\{\mbox{Since } F=F_{MAX}^{EK}\}$\\

$\{A, B, S\}$\\

Then, we have:

\begin{equation}
{\Upsilon}_{p,F}(U,\{U.\{A.V\}_{k_{bs}}\}_{k_{bs}})=\{A, B, S\}
\label{eq8}
\end{equation}

2- $\forall V$:\\

a- Receiving step: $R_{S^{i}}^-=\{A.U.\{B.V\}_{k_{as}}\}_{k_{bs}}$ \textit{(when receiving, we use the upper bound)}\\

\begin{tabular}{lll}
   $\!\!\!\!\!\!\!\!F'(V,\{A.U.\{B.V\}_{k_{as}}\}_{k_{bs}})$ & $\!\!\!\!\!\!=$ & $\!\!\!\!\!\!\!F(V,\partial  [\overline{V}]\{A.U.\{B.V\}_{k_{as}}\}_{k_{bs}})$  \\
   & $\!\!\!\!\!\!=$ & $\!\!\!\!\!\!F(V,\{A.\{B.V\}_{k_{as}}\}_{k_{bs}} )$ \\
    & $\!\!\!\!\!\!=$  & $\!\!\!\!\!\!F(V,\partial  [\overline{V}]\{A.U.\{B.V\}_{k_{as}}\}_{k_{bs}})$\\
    & $\!\!\!\!\!\!=$  &$\!\!\!\!\!\!F(V,\{A.\{B.V\}_{k_{as}}\}_{k_{bs}} )$ \\
\end{tabular}
$=\left\{
    \begin{array}{ll}
        \{A, B, S\} & \mbox{if } k_{as} \mbox{ is the external protective key } \mbox{of } V \mbox{ in}\\
        &~~\mbox{ the received message } \{A.\{\textcolor{magenta}{B}.V\}_{k_{\textcolor{magenta}{as}}}\}_{k_{bs}} \\
                          \\
        \{A, B, S\} & \mbox{if } k_{bs} \mbox{ is the external protective key } \mbox{of } V \mbox{ in}\\
        &~~\mbox{ the received message } \{\textcolor{magenta}{A}.\{\textcolor{magenta}{B}.V\}_{k_{as}}\}_{k_{\textcolor{magenta}{bs}}} 
   \end{array} \right.$\\$ $\\
\begin{tabular}{lll}
   ~~~~~~~~~~~~~~~~~~~~~~~~~~~~~~~~~~& $\!\!\!\!\!\!=$  & $\{A, B, S\}$
\end{tabular}   

 
Then, we have:

\begin{equation}
F'(V, \{A.U.\{B.V\}_{k_{as}}\}_{k_{bs}})=\{A, B, S\}
\label{eq9}
\end{equation}

b- Sending step: $r_{S^{i}}^+=\{U.\{A.V\}_{k_{bs}}\}_{k_{bs}}$ \textit{(when sending , we use the lower bound)}\\

 $\forall V.\{(m',\sigma') \in \tilde{\cal{M}}_{p}^{\cal{G}}\otimes\Gamma|{m'\sigma' = r_{S^{i}}^+\sigma' } \}$\\
 
 $=\forall V.\{(m',\sigma') \in \tilde{\cal{M}}_{p}^{\cal{G}}\otimes\Gamma|{m'\sigma' = \{U.\{A.V\}_{k_{bs}}\}_{k_{bs}} \sigma' } \}$ \\
 
 $=\{(\{\{U_2.\{A_7.V_2\}_{K_{B_{6}S_5}}\}_{K_{B_{6}S_5}},\sigma_1'),$
 
 ~~~~$(\{N_{B_{4}}^i.\{A_5.Z_1\}_{K_{B_{4}S_3}}\}_{K_{B_{4}S_3}},\sigma_2')\}$ such that:

$$\left\{
    \begin{array}{l}
      \sigma_1'=\{ U_2 \longmapsto U, A_7\longmapsto A, V_2 \longmapsto V, {K_{B_{6}S_5}}\longmapsto  {k_{bs}}\}   \\
      \sigma_2'=\{ U \longmapsto N_{B_{4}}^i, A_5\longmapsto A, Z_1 \longmapsto V, {K_{B_{4}S_3}}\longmapsto  {k_{bs}}\}  \\
    \end{array}
\right.$$

${\Upsilon}_{p,F}(V,\{U.\{A.V\}_{k_{bs}}\}_{k_{bs}})$\\

$=\{\mbox{Definition of the lower bound of the witness function}\}$\\
$$F'(V,\{U_2.\{A_7.V_2\}_{K_{B_{6}S_5}}\}_{K_{B_{6}S_5}} \sigma_{1}')$$ $$\sqcap$$ 
$$F'(V,\{N_{B_{4}}^i.\{A_5.Z_1\}_{K_{B_{4}S_3}}\}_{K_{B_{4}S_3}} \sigma_{2}') $$

$=\{\mbox{Setting the static neighborhood}\}$
$$F'(V,\{U.\{A.V_2\}_{k_{bs}}\}_{k_{bs}} \sigma_{1}') \sqcap F'(V, \{ N_{b}^i.\{A.V_2\}_{k_{bs}}\}_{k_{bs}} \sigma_{2}')$$ 

$=\{\mbox{Definition } \ref{Fder}\}$
$$F(V_2,\partial[\overline{V_2}] \{U.\{A.V_2\}_{k_{bs}}\}_{k_{bs}}) \sqcap F(V_2,\partial[\overline{V_2}] \{N_{b}^i.\{A.V_2\}_{k_{bs}}\}_{k_{bs}})$$

$=\{\mbox{Derivation in Definition } \ref{derivation}\}$
$$F(V_2,\{\{A.V_2\}_{k_{bs}}\}_{k_{bs}})\sqcap F(V_2,\{N_{b}^i.\{A.V_2\}_{k_{bs}}\}_{k_{bs}})$$ 

$=\{\mbox{Since } F=F_{MAX}^{EK}\}$\\

$\{ A, B, S\}$\\

Then, we have:

\begin{equation}
{\Upsilon}_{p,F}(V,\{U.\{A.V\}_{k_{bs}}\}_{k_{bs}})=\{A, B, S\}
\label{eq10}
\end{equation}

3- Conformity  with Lemma \ref{PAT}:\\

From (\ref{eq7}) and (\ref{eq8}), we have:\\
\begin{equation}
{\Upsilon}_{p,F}(U,\{U.\{A.V\}_{k_{bs}}\}_{k_{bs}}) \sqsupseteq \ulcorner U \urcorner \sqcap F'(U,\{A.U.\{B.V\}_{k_{as}}\}_{k_{bs}} )
\label{eq11}
\end{equation}

From (\ref{eq9}) and (\ref{eq10}), we have:\\
\begin{equation}
{\Upsilon}_{p,F}(V,\{U.\{A.V\}_{k_{bs}}\}_{k_{bs}}) \sqsupseteq \ulcorner V \urcorner \sqcap F'(V,\{A.U.\{B.V\}_{k_{as}}\}_{k_{bs}}  
\label{eq12}
\end{equation}

From (\ref{eq11})  and (\ref{eq12}),  we have:  the generalized role of $S$ respects Lemma \ref{PAT}. ~~~~~~~~~~~~~~~~~~~~~~~~~~~~~~~~~~~~~~~~~~~~~~~~(III)$ $\\

From (I) and (II) and (III), we conclude that: $p$ respects Lemma~\ref{PAT}. ~~~~~~~~~~~~~~ ~~~~~~~~~~~~~~~~~~~~~~~~~~~~~~~~~ ~~~~~(IV)\\
$ $\\
Now that the first condition of Theorem \ref{PAT2} is satisfied (i.e. the protocol is proven increasing, so correct for secrecy), let us examine the second one. Indeed, the message received by the authenticator $B$ to authenticatee  $A$ is $\{N_b^i.\{A.Z\}_{k_{bs}}\}_{k_{bs}}$. The challenge to be verified by $B$ is the nonce $N_{b}^{i}$. We have:\\ $ $\\
\begin{tabular}{lll}
   $\!\!\!\!\!F'(N_{b}^{i}, \{N_b^i.\{A.Z\}_{k_{bs}}\}_{k_{bs}}$ & $\!\!\!\!\!=$ &  $\!\!\!F(N_{b}^{i}, \partial[\overline{N_{b}^{i}}] \{N_b^i.\{\textcolor{magenta}{A}.Z\}_{k_{bs}}\}_{k_{bs}}$ \\
 & $\!\!\!\!\!=$ & $\!\!\!F(N_{b}^{i}, \{N_b^i.\{\textcolor{magenta}{A}\}_{k_{bs}}\}_{k_{bs}})$ \\
    & $\!\!\!\!\!=$ & $\!\!\!\{\textcolor{magenta}{A},B,S\}$ \\
 \end{tabular}
 $ $\\
Then, we can easily see that: 
$$A\in F'(N_{b}^{i}, \{N_b^i.\{A.Z\}_{k_{bs}}\}_{k_{bs}})$$ $$\wedge$$  $$F'(N_{b}^{i}, \{N_b^i.\{A.Z\}_{k_{bs}}\}_{k_{bs}})  \sqsupset \bot~~~\mbox{(V)}$$ 
 $ $\\
From (IV) and (V) and Theorem \ref{PAT2}, we can declare now, and only now, that: {The modified version of the Woo-Lam protocol is correct with respect to authentication.}

\section{Comparison with similar works}

In the  witness functions logic, to prove authentication, a protocol must be verified for secrecy first. For that, the upper bound of a witness function bases  its calculation on safe identities collected from a piece of message that is fully invariant by opponent in order to deprives this latter from using his capabilities to forge these identities. Besides, in every single sending step of the protocol, the lower bound makes sure that the protocol does not endow the opponent of new rules  that could mislead a regular agent by using suspicious messages that '\textit{'look like}'' a regular one but,  in fact, have been gathered from previous sessions. To do so, this bound considers all encryption patterns that might be sources of a final trace and verifies whether or not a suspicious identity could be inseminated in some variable when the protocol is running. If this is the case, the analysis is immediately aborted and no result for the protocol correctness is given with respect to secrecy. This way we  treat secrecy leads to an interesting observation: ''If a protocol is correct for secrecy, then any atomic messages arrives to its destination in a safe state''. That means that  its evaluation environment, based on a protective  key and all the identities beyond, does not contain any suspicious identity. All of them are reliable. It turns out that the authenticator has just to  make sure that the identity of the authenticatee  is present, witnessing finally the origin of the authenticating message. This is made available by the upper bound that is called, once again, to provide this authentication service, too. In literature, we can point out an interesting work which is similar in some aspects to ours. It is the work of Houmani published in ~\cite{Houmani1,Houmani3,Houmani8,Houmani5}. In this work, she defined two functions to estimate the level of security of atoms in messages called respectively $\mbox{DEK}$ and $\mbox{DEKAN}$. The $\mbox{DEK}$ function is based on the direct key only whereas  $\mbox{DEKAN}$ is based on the direct key and neighbors. The $\mbox{DEK}$ function  is very limited in practice. However, the main drawback of the $\mbox{DEKAN}$ function  is that it is not variable free. For instance, $\mbox{DEKAN}(\alpha,\{\alpha.A.X\}_{k_{bs}})=\{A,,B,S\} \sqcap \ulcorner X \urcorner$, where $X$ is a variable having an unknown level of security $\ulcorner X \urcorner$ in the context. The fact that a function has outputs with variables sets up a real barrier when comparing two security levels, especially when we have more than one variable in the same message. As a result, very few protocols have been shown correct with respect to secrecy and, as far as we know, no further researches have been published with respect to authentication, very likely because of variables in output. With witness functions, we do not have this problem owing to derivation used in the witness functions'  bounds that eliminate variables in output. Hence, the comparison between security levels is made possible and easier.  Several tools have been proposed in the state-of-the-art of cryptographic protocols to verify authentication in security protocols. Among them, we can cite an interesting one: the AVISPA tool~\cite{TrustComVigano200661}. This tool consists of an aggregation of three model checkers and a tree automata verifier. These four components are:
\begin{itemize}
\item the Constraint-Logic-based Attack Searcher:  applies constraint solving to run both protocol falsification and verification for bounded sessions; 

\item the On-the-fly Model-Checker:  it carries out  a protocol verification by exploring the transition system. It considers both typed and untyped protocol models;

\item the SAT-based Model-Checker: it uses typed protocols and carries out bounded session verification  by calling a SAT solvers to reduce problem input;

\item the TA4SP: is a tree automata that carries out unbounded protocol verification by estimating the opponent knowledge using rewriting techniques.\\

\end{itemize}

The AVISPA tool was successful in finding new  attacks, for instance,  on the ISO-PK3 protocol and the IKEv2 protocol with digital signatures (a man-in-the-middle attack). However, the main drawback of model checkers remains the state explosion problem that we do not face using our approach. Another interesting tool is the ProVerif one~\cite{TrustCOMChevalBlanchetPOST,TrustCOMBlanchetBook}.  This latter uses an abstract representation of a protocol by Horn clauses and supports several cryptographic primitives described by rewrite rules or equations. The main limitation of this kind of verifiers is the halting problem where a program could never end. That is why binding sessions becomes usually a must. In our approach, we are not concerned about this problem since our approach is fully static.

\section{Conclusion and future work}

In this paper, we strongly believe that we are giving a new analytic direction for protocols' verification against authentication flaws by using our recent finding: the witness functions.  In a future work, we will  experiment our approach on more protocols in order to know how wide is the range of protocols that could be analyzed with it and how reasonable is the rate of false negatives. A special attention will be paid to protocols that do not run in isolation in order to prevent multi-protocol attacks~\cite{TrustComArapinisCD15,TrustComCremers2012}. Finally, anonymity is one of our immediate targets. In fact, this property seems to be the opposite of authentication in the sense that the link between the sender identity and the secret must be shown broken during  the communication to make sure that an opponent cannot reconstruct it. This is a major concern in e-voting protocols, for example. In theory, witness functions should be useful to prove this property  by setting sufficient conditions for anonymity.



\bibliographystyle{ieeetr}
\bibliography{Ma_these}

\section*{ Appendix: Notations}
\footnotesize{
\begin{itemize}
\item[+] We denote by ${\cal{C}}=\langle{\cal{M}},\xi,\models,{\cal{K}},{\cal{L}}^\sqsupseteq,\ulcorner.\urcorner\rangle$ the context of verification containing the parameters that affect the analysis of a protocol:
\begin{itemize}
\item[$\bullet$] ${\cal{M}}$: is a set of messages built from the algebraic signature $\langle\cal{N}$,$\Sigma\rangle$ where ${\cal{N}}$ is a set of atomic names (nonces, keys, principals, etc.) and $\Sigma$ is a set of functions ($enc$:\!: encryption\index{Encryption}, $dec$:\!: decryption\index{Décryption}, $pair$:\!: concatenation (denoted by "." ), etc.). i.e. ${\cal{M}}=T_{\langle{\cal{N}},\Sigma\rangle}({\cal{X}})$. We use $\Gamma$ to denote the set of all substitution from $ {\cal{X}} \rightarrow {\cal{M}}$.
We denote by $\cal{A}$ all atomic messages (atoms)  in ${\cal{M}},$ by ${\cal{A}}(m)$ the set of atomic messages in $m$ and by ${\cal{I}}$ the set of principals including the opponent $I$. We denote by $k{^{-1}}$ the reverse form of a key $k$ and we assume that $({k^{-1}})^{-1}=k$.
\item[$\bullet$] $\xi$: is the theory that describes the algebraic properties of the functions in $\Sigma$ by equations (e.g. $dec(enc(x,y),y^{-1})=x$). 
\item[$\bullet$] $\models$: is the inference system of the opponent under the theory. Let $M$ be a set of messages and $m$ a message. $M$ $\models$ $m$ means that the opponent can infer $m$ from $M$ using his capabilities. We extend this notation to traces as follows: $\rho$ $\models$ $m$ means that the opponent can infer $m$ from the trace $\rho$.
\item[$\bullet$] ${\cal{K}}$ : is a function from ${\cal{I}}$ to ${\cal{M}}$, that assigns to any principal a set of atomic messages describing his initial knowledge. We denote by $K_{{\cal{C}}}(I)$ the initial knowledge of the opponent, or simply $K(I)$ where the context of verification is obvious.
\item[$\bullet$] ${\cal{L}}^\sqsupseteq$ : is the security lattice $({\cal{L}},\sqsupseteq, \sqcup, \sqcap, \bot,\top)$ used to assign security values to messages. 
A concrete example of a lattice is $ (2^{\cal{I}},\subseteq,\cap,\cup,\cal{I}, \emptyset)$ that will be used in this paper. 
\item[$\bullet$] $\ulcorner .\urcorner$ : is a partial function that assigns a value of security (type) to a message in ${\cal{M}}$. Let $M$ be a set of messages and $m$ a single message. We write $\ulcorner M \urcorner \sqsupseteq \ulcorner m \urcorner$ when
$\exists m' \in M. \ulcorner m' \urcorner \sqsupseteq \ulcorner m \urcorner$
\end{itemize}
\item[+] Let $p$ be a protocol, we denote by $R_G(p)$ the set of the generalized roles extracted from $p$. A generalized role is an abstraction of the protocol where the emphasis is put on a specific principal and all the unknown messages are replaced by variables. More details about the role-based specification could be found in~\cite{Debbabi11}.
We denote by ${\cal{M}}_p^{\cal{G}}$ the set of messages generated by $R_G(p)$, by ${\cal{M}}_p$ the set of closed messages (ground terms) generated by substitution in terms in ${\cal{M}}_p^{\cal{G}}$. We denote by $R^-$ (respectively $R^+$) the set of received messages (respectively sent messages) by a principal in the role $R$. Conventionally, we use uppercases for sets or sequences and lowercases for single elements. For example $M$ denotes a set of messages, $m$ a message, $R$ a role composed of sequence of steps, $r$ a step and $R.r$ the role ending by the step $r$. A valid trace is a ground term obtained by substitution in the generalized roles. We assume that the opponent has the full-control of the net and is able to defeat any operation $f$ in $\Sigma$ except encryption if he does not have the decryption key, as described in the Dolev-Yao model~\cite{DolevY1}. 
\end{itemize}
}
\end{document}